# Reconstruction-based spectroscopy using CMOS image sensors with random photon-trapping nanostructure per sensor

**Ahasan Ahamed[1*], Cesar Bartolo-Perez[1], Ahmed Sulaiman Mayet[1], Soroush Ghandiparsi[1], Lisa McPhillips[1], Shih-Yuan Wang[2], and M. Saif Islam[1]**

[1]Electrical and Computer Engineering Department, University of California – Davis, Davis, California, USA, 95616
[2]W&WSens Devices, Inc., 4546 El Camino, Suite 215, Los Altos, California, USA, 94022

## ABSTRACT

Optical spectrometers are widely used scientific equipment with many applications involving material characterization, chemical analysis, disease diagnostics, surveillance, etc. Emerging applications in biomedical and communication fields have boosted the research in the miniaturization of spectrometers. Recently, reconstruction-based spectrometers have gained popularity for their compact size, easy maneuverability, and versatile utilities. These devices exploit the superior computational capabilities of recent computers to reconstruct hyperspectral images using detectors with distinct responsivity to different wavelengths. In this paper, we propose a CMOS compatible reconstruction-based on-chip spectrometer pixels capable of spectrally resolving the visible spectrum with 1 nm spectral resolution maintaining high accuracy (>95 %) and low footprint (8 μm × 8 μm), all without the use of any additional filters. A single spectrometer pixel is formed by an array of silicon photodiodes, each having a distinct absorption spectrum due to their integrated nanostructures, this allows us to computationally reconstruct the hyperspectral image. To achieve distinct responsivity, we utilize random photon-trapping nanostructures per photodiode with different dimensions and shapes that modify the coupling of light at different wavelengths. This also reduces the spectrometer pixel footprint (comparable to conventional camera pixels), thus improving spatial resolution. Moreover, deep trench isolation (DTI) reduces the crosstalk between adjacent photodiodes. This miniaturized spectrometer can be utilized for real-time in-situ biomedical applications such as Fluorescence Lifetime Imaging Microscopy (FLIM), pulse oximetry, disease diagnostics, and surgical guidance.

**Keywords:** Spectroscopy, spectrometer-on-a-chip, reconstruction-based, photon-trapping nanostructures, CMOS compatible, miniaturization, fluorescent lifetime imaging microscopy (FLIM), hyperspectral imaging.

## 1. INTRODUCTION

A spectrometer is a powerful optical instrument that can separate a mixed wavelength of light into individual narrow bands of color or spectrum. They are widely used in both scientific research and practical applications involving food processing, chemical imaging, molecular analysis, mineralogy, disease diagnostics, etc[1,2]. The state-of-the-art spectrometers use bulky optical elements that restrict their uses in emerging biomedical applications such as surgical guidance and in-situ disease diagnosis. For the past few decades, researchers have focused on miniaturizing spectrometers using dispersive ray-optics[3,4], tunable narrowband filters[5,6], and Mach-Zehnder interferometry[7,8].

Recently, the surge of computational power with significant reductions in microprocessor size and cost has led to the development of reconstruction-based spectroscopy techniques. With the help of modern processing power, we can essentially make a spectrometer-on-a-chip using just a set of unique photodetectors[9,10]. Reconstructive hyperspectral imaging can be done by a set of unique photodiodes created by engineering their spectral response by modifying their bandgap[11] or by the addition of photonic crystal slabs on top of a detector array[12]. However, the fabrication technique for an engineered bandgap is challenging and cannot be reliably reproduced. The addition of a photonic crystal slab is an attractive solution because of the potential for CMOS integration; however, it affects the efficiency of the detectors that are crucial for biomedical applications as they operate with minimal light power. Higher light intensities may burn the biological sample and may cause irreversible damage to the patient.

[*]aahamed@ucdavis.edu; phone 530-5742762

We propose a miniaturized spectrometer-on-a-chip with high spectral resolution (1 nm), low footprint (8 μm × 8 μm), and decent accuracy (>95%) over the visible range (400 nm to 700 nm) for biomedical applications for in-situ imaging and surgical guidance. This spectrometer is equipped with an array of 4 × 4 silicon photodiodes with random nanostructures integrated into them. Each photodiode is separated by a thin layer of oxide acting as Deep Trench Isolation (DTI) which reduces the crosstalk between the photodiodes. These devices are also CMOS compatible making them a reliable solution for mass production. The low footprint would allow higher spatial resolution useful for precisely finding malignant tumor cells and tumor diagnosis, while the broad range covered by the spectrometer allows different fluorescence emitters to be used in Fluorescent Lifetime Imaging Microscopy (FLIM). This spectrometer is an optimal solution for emerging biomedical applications such as cancer diagnosis, heart, and circulatory pathology, FLIM, etc[13–15]. Real-time applications such as surgical guidance for breast cancer, gall bladder stone, renal tumor require less sophisticated yet effective spectrometers that can substantially impact the clinical treatment. Such a miniaturized spectrometer array can easily be placed on the tip of a surgical knife and can scan through the patient's malignant tumor cells in real-time.

## 2. CONCEPTS

### 2.1 Reconstruction-based Spectroscopy

In a traditional spectrometer, we use diffractive optics to disperse the light spatially and collect them in a detector array to read out the relative amplitude of different wavelengths. The major disadvantage for these spectrometers is that they require a long optical path for the propagation of light. Then, to evade this situation, we started to use narrowband filters on top of the detector array that would limit the amount of light that can activate the detectors. This worked fine for a limited number of channels, in other words, the spectral resolution deteriorated significantly. With the uprising of modern computational power, reconstruction-based spectroscopy became quickly popular because it circumvents the necessity of large optical instruments and also can provide better spectral resolution. In a reconstruction-based spectrometer, we use a set of detectors equipped with capabilities to differentiate light wavelengths, that is, they would respond differently with different wavelengths of light. They are then trained with a large variety of input signals and when an unknown light signal is illuminated on them, the improved computational capabilities can identify the spectral information by carefully observing the behavior of the different output levels of the detectors.

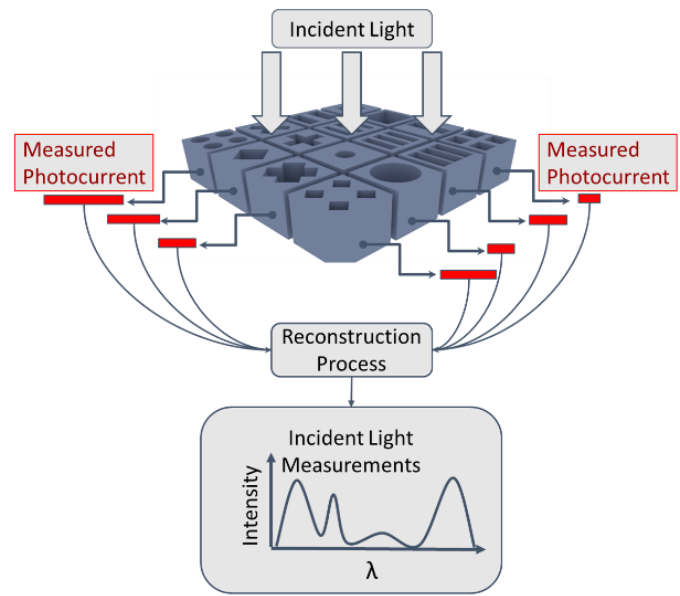

Figure 1: A schematic of reconstruction-based spectroscopy with a spectrometer-on-a-chip formed by 4 × 4 silicon photodiodes with random nanostructures integrated in them.

### 2.2 Device Structure

The spectrometer-on-a-chip is formed by an array of small silicon photodiodes, each having an active area with dimensions 2 μm × 2 μm, and a depth of 1.5 μm as shown in figure 1. The photodiode sits on top of a thick layer of silicon-di-oxide working as the back reflector and reducing the transmittance of the incoming light. Each photodiode is separated by a 100 nm layer of oxide that acts as the Deep Trench Isolation (DTI). This helps in reducing the crosstalk between each photodiode. Each photodiode is made unique by the integration of random nanostructures that modulate the light-matter interaction in the active area. This allows us to have unique responsivity from each of the detectors which can then be utilized to reconstruct the hyperspectral images by computational methods.

## 2.3 Photon-trapping Nanostructures

Photon-trapping nanostructures have been used in the scientific community to improve the quantum efficiency of photodiodes and solar cells for decades[16,17]. These nanostructures are basically integrated nanoholes or nanopillars embedded on top of the active layer of the detectors that can modulate the light-matter interaction. These subwavelength structures can interact with light and change their propagation direction from a vertical orientation to a lateral orientation by allowing only lateral propagation modes to exist in the semiconductor. Since light interacts with mostly subwavelength structures, we can design nanostructures with distinct sizes and shapes such that they would interact differently with different wavelengths. Thus, this allows us to create photodiodes that would have unique responsivity to light over a specific wavelength range.

# 3. METHODS AND PROCEDURE

## 3.1 Fabrication of Random Photon-trapping Nanostructures

In this paper, the proposed detector-only spectrometer is constituted of an array of p-i-n photodiodes with integrated random photon-trapping nanostructures. These nanostructures can be fabricated into the photodiodes by etching the surface of the detector through the i-layer. They can be etched on silicon with a high aspect ratio by alternating cycles of Deep Reactive Ion Etching (DRIE) and passivation. Details on the fabrication techniques can be found in the reference[18]. The different shapes and sizes of the nanostructures used in this paper are shown in figure 1. Each of the photodiodes is separated by DTI that blocks light interference between the photodiodes. The signals are collected from each of the photodiodes individually by top and bottom contacts connected internally to a transmission line for the array.

## 3.2 Training

For reconstruction-based spectroscopy to work we first need to characterize our photodiodes and learn their behaviors in response to different illumination conditions. So firstly, we obtain their quantum efficiency over the visible spectrum from 400 nm to 700 nm wavelength range. Since this is a simulation paper, we obtained the data by simulating the devices in the Lumerical FDTD solutions and obtaining their absorption profile over the given wavelength range. Next, using MATLAB we illuminate the detectors with lights with varying peak wavelengths with different breadths and record the estimated photocurrent in the detectors. The estimated photocurrent is derived from the quantum efficiency of the photodiodes in the wavelength range and the input light spectrum. For the training sample, the peak amplitudes varied from 400 nm to 700 nm, and the breadths are measured in Full-Width-Half-Maximum (FWHM) varying from 40 nm to 120 nm. These training data were selected as we are focusing on biomedical applications which deal with spectrally broader signals in the visible wavelength spectrum.

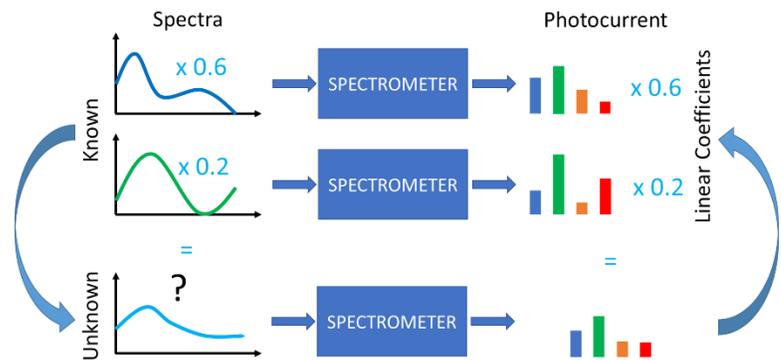

## 3.3 Reconstruction

When the light of an unknown spectrum is illuminated onto the reconstruction-based spectrometer, the different photodiodes generate different photocurrent amplitudes that are recorded by the device. These photocurrents are then reconstructed using the known photocurrent outputs that were previously recorded during the training procedure of the device. So, we calculate the linear coefficient of the known photocurrent samples by matrix multiplication. Then, these linear coefficients are used to estimate the unknown input light spectrum by multiplying them to the known light spectra. The process is illustrated in detail in figure 2.

Figure 2: Demonstration of reconstruction algorithm for our detector-only spectrometer. We train our spectrometer with known spectra and record their respective photocurrent. The output photocurrent from an unknown spectrum is used to find the linear coefficients of the known photocurrent using matrix multiplication techniques. The same coefficients are then used to estimate the unknown spectrum by multiplying the known spectra with the linear coefficients.

## 4. RESULTS AND DISCUSSION

### 4.1 Quantum Efficiencies of the Photodiodes with Random Nanostructures

We use Lumerical FDTD solutions to generate the quantum efficiencies of the photodiodes with different nanostructures as shown in figure 3. Here the quantum efficiencies are demonstrated for the respective nanostructured photodiodes over a wavelength range from 400 nm to 700 nm. The top view of the respective nanostructured photodiode is shown in the inset of each subfigure.

The results show that the photodiodes with a single nanostructure can also modulate the light propagation and thus impact their quantum efficiencies. Most of the profiles have a similar pattern arising from the intrinsic absorption characteristics of silicon. However, the nanostructures contribute to peaks and valleys at different wavelengths which help to create the distinctive features of the photodiodes necessary for reconstruction-based spectroscopy. The more uniqueness in these profiles generally contributes to a better performance in the reconstruction-based spectrometer.

One important thing to note is that the maximum quantum efficiency achieved in such photodiodes is about 80% over the wavelength range from 400 nm to 700 nm. It is expected as the rest of the light is mostly reflected from the top surface of the detector. The transmittance is low for shorter wavelengths, and it increases with increasing wavelength as observed in any conventional silicon photodiode. This explains the quantum efficiency dropping off at longer wavelengths.

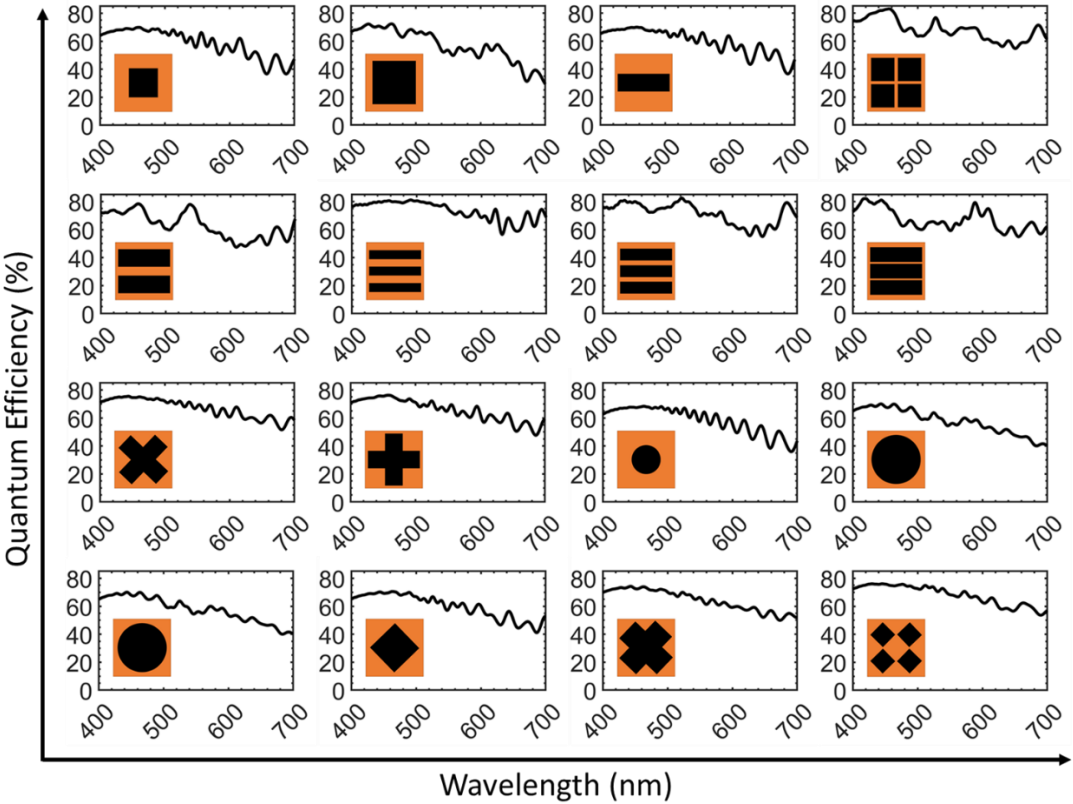

Figure 3: The detector-only spectrometer is comprised of 4×4 photodiode arrays. The quantum efficiencies and the top view of each nanostructured photodiodes are shown here in the subfigures. We observe variation in the quantum efficiencies over the visible spectrum from 400 nm to 700 nm wavelength due to the random nanostructures integrated in the photodiodes.

## 4.2 Reconstructed Spectra

We used the quantum efficiencies shown in figure 3 and used them to train the system to learn about our photodiodes. Then we generated several random spectra between 40 nm to 120 nm FWHM and reconstructed them with 1 nm spectral resolution using our reconstruction algorithm discussed earlier. Some of these results are shown in figure 4. Here, the normalized intensity of input light is shown in a solid blue line, and the reconstructed spectrum is shown in a dotted red line with square markers. The original and reconstructed colors are also shown at the bottom of each figure. The Root Mean Square Error (RMSE) shows that the spectrometer can successfully reconstruct wavelengths within the spectral width range with more than 95% accuracy.

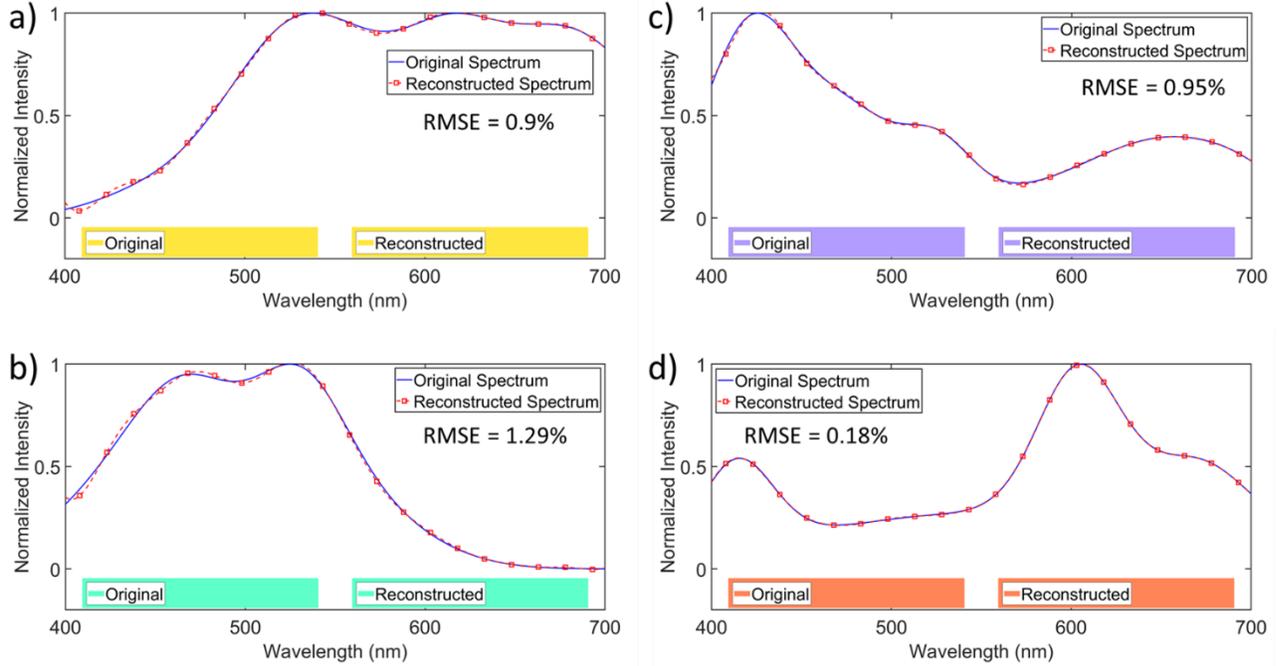

Figure 4: (a-d) The reconstructed spectrum (red dotted line with square marker) is compared with the original spectrum (blue solid line) over the wavelength range from 400 nm to 700 nm. The original and reconstructed colors are shown in the inset of respective figures. We observe Root Mean Square Error (RMSE) of 0.9%, 1.29%, 0.95%, and 0.18% for the reconstructed spectra respectively.

## 4.3 Root Mean Square Error in Reconstruction

The RMSE is calculated by taking the Root Mean Square (RMS) of the error between the reconstructed spectrum from the original input spectrum at every wavelength point (with 1 nm spectral resolution). RMSE was chosen as it is a good measure of the standard deviation from the original signal. From figure 5 we observe that the RMSE is lower than 2%. This is due to the distinct responsivity of the photodiodes and better reconstruction algorithm. We simulated over 10,000 input spectra and made a histogram chart of the RMS distribution as shown in figure 5. We observe that the RMSE is distributed with a mean of 1.17% average error and is mostly contained within 2.5% error.

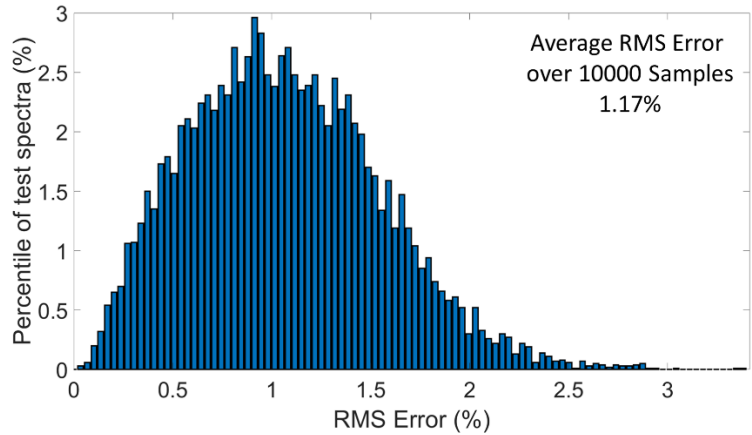

Figure 5: RMS error distribution for 10000 sample reconstructed spectra with an average RMS error of 1.17%.

We also simulated light wavelengths with varying peaks from 400 nm to 700 nm wavelengths for different FWHM to observe the behavior of the reconstruction algorithm over the spectral range. Figure 6 shows the average RMSE observed for different FWHM from 40 nm to 120 nm, and the profile over the visible spectral range. We observe that the spectrometer can easily reconstruct input spectra with FWHM from 40 nm to 120 nm with less than 5% RMSE. We also show in the inset that the reconstruction of the input spectrum with peak 540 nm, FWHM 40 nm shows RMSE of 4.98% while for input spectrum with peak 620 nm, FWHM 120 nm shows RMSE of only 0.38%. The error profile as shown in figure 6 is generated by plotting the RMSE for individual spectrum with respective peak and FWHM.

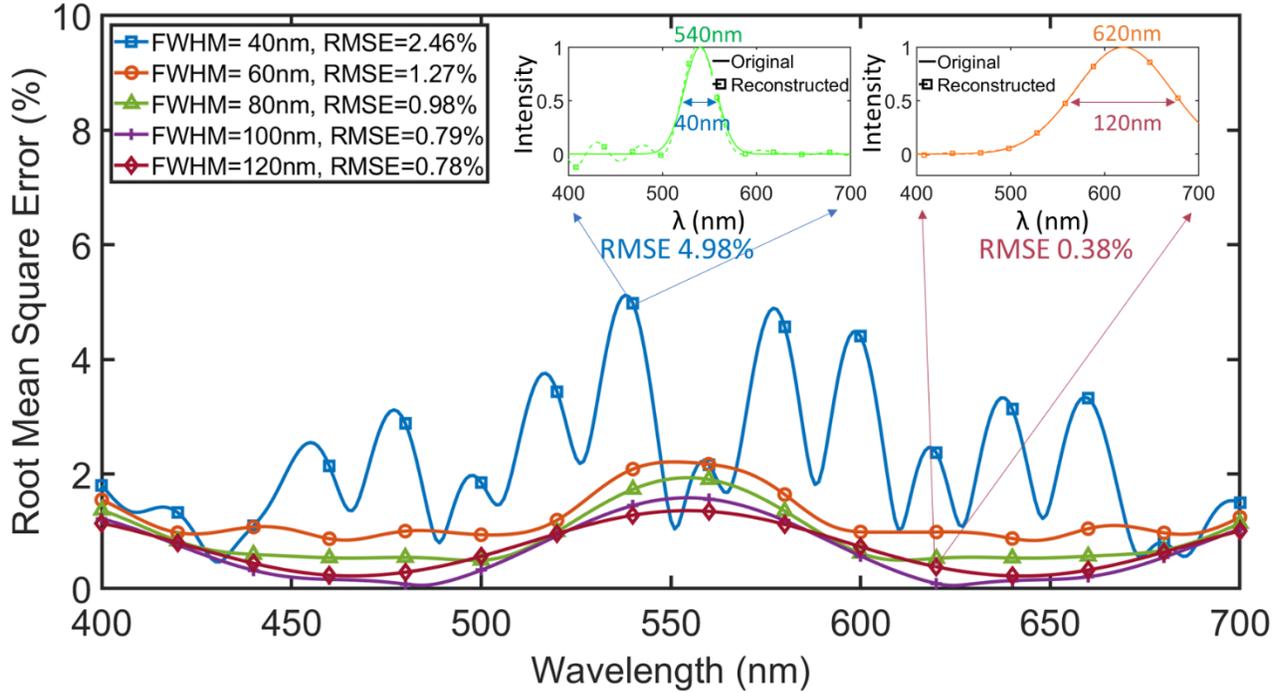

Figure 6: The RMS error for input spectra with respective peak wavelength and spectral width is plotted over the wavelength range from 400 nm to 700 nm. The average RMS error for spectral widths from 40 nm to 120 nm is shown on top left corner. Two points in the graph are illustrated by showing the reconstructed spectrum for 540 nm peak wavelength with 40 nm FWHM, and 620 nm peak wavelength with 120 nm FWHM. They show RMSE of 4.98% and 0.38% respectively. The results show that the spectrometer can reliably operate within the operating spectral width with more than 95% accuracy.

We observe in figure 6 that for FWHM from 60 nm to 120 nm, the reconstruction error is less than 2%. However, for FWHM 40 nm, the error increases up to 5%. As we narrow down the spectral width the RMSE increases rapidly. This is because of the lack of uniqueness generated by the nanostructures in the quantum efficiency profile. Since the nanostructures are large compared to the spectral width, it interacts with a broader range of wavelengths compared to the narrow spectral width. Therefore, we cannot reconstruct sharp input spectra with narrow spectral widths. However, for our application of interest, most biomedical fluorescence emitters emit with a broad spectral width from around 50 nm to more than 100 nm. Therefore, our spectrometer can be reliably used for these applications.

## 5. CONCLUSION

The miniaturized spectrometer-on-a-chip proposed in this paper would revolutionize the use of spectrometers in *in-situ* and *in-vivo* biomedical applications. The proposed reconstruction-based spectrometer circumvents the use of bulky optical parts, narrowband filters, or any additional elements. The use of random nanostructures generates unique responsivity patterns in the detectors. With the help of advanced computational power, we can create a detector-only spectrometer array that can be easily mounted on a surgical knife and can be used for real-time surgical guidance. This device provides spectral information about the sample with a high spatial and spectral resolution while maintaining a reliable reconstruction accuracy above 95% for input spectral width from 40 nm to 120 nm FWHM over the visible spectrum. This covers most of the fluorescent emitters used in fluorescence lifetime imaging microscopy (FLIM) making them suitable for this imaging modality. We observe that the RMS error lies within 2.5% for over 10000 random samples making them reliable for repeated use. Also, the fabrication process of these devices is CMOS compatible, therefore, reliable, repeatable reproduction of miniaturized spectrometers-on-a-chip are possible. We believe these spectrometer arrays will find their application in FLIM, disease diagnosis, tumor boundary detection, breast cancer surgery, heart pathology, etc.

We are researching on increasing the spectral range of the spectrometers into the near-infrared region for accommodating more applications such as pulse oximetry and fetal heart rate. More research on the nanostructures is necessary to improve the reconstruction accuracy for narrow spectral widths.

## ACKNOWLEDGEMENTS

This work was supported in part by the Dean's Collaborative Research Award (DECOR) of UC Davis College of Engineering, the S. P. Wang and S. Y. Wang Partnership, Los Altos, CA. Cesar Bartolo-Perez acknowledges the National Council of Science and Technology (CONACYT) and UC-MEXUS for the Doctoral fellowship.